\documentclass[aps,prc,showpacs,superscriptaddress,preprint,floatfix]{revtex4-1}

\usepackage[dvips]{graphicx}
\usepackage{color}
\usepackage{amssymb}
\usepackage{amsmath}
\usepackage{braket}
\usepackage{hyperref}
\usepackage{bm}
\usepackage{multirow}
\newcommand{\nc}{\newcommand}           % new command
\nc{\vc}[1]     {\mbox{\boldmath $#1$}} % boldmath(vector)
\nc{\mapleft}[1]{                       % something under arrow
 \smash{\mathop{                        %
  \hbox to 0.90cm{\rightarrowfill} }\limits_{#1}}}
\nc{\figwidth}{0.8}                    % figure width, in unit of textwidth, preprint
\nc{\mydraft}	{\setlength{\topmargin}{-1.5cm}}
\mydraft   

\begin{document}
\title{The study of the $\alpha$ formation probability in $^{10}$Be and $^{12}$Be within the microscopic cluster model}

\author{Qing Zhao} \email[]{zhaoqing91@zjhu.edu.cn}
\affiliation{School of Science, Huzhou University, Huzhou 313000, Zhejiang, China}
\author{Masaaki Kimura}
\affiliation{Nuclear Reaction Data Centre (JCPRG), Hokkaido University, Sapporo 060-0810, Japan}
\affiliation{Department of Physics, Hokkaido University, Sapporo 060-0810, Japan}
\affiliation{RIKEN Nishina Center, Wako, Saitama 351-0198, Japan}
\author{Bo Zhou}
\affiliation{Institute of Modern Physics, Fudan University, Shanghai 200433, China}
\author{Seung-heon Shin}
\affiliation{Department of Physics, Hokkaido University, Sapporo 060-0810, Japan}

% cSpell: enable

\begin{abstract}
The $\alpha$ clustering in the $^{10}$Be and $^{12}$Be has been studied within the framework of the real-time evolution method (REM). By using the effective interaction tuned to reproduce the charge radii and the threshold energies, we have evaluated the $\alpha$ reduced width amplitude (RWA) and spectroscopic factor (S-factor) for the ground and excited states. With several improvements made in this work, comparing with our previous calculations, larger rate of clustering results are obtained for the $^{10}$Be and $^{12}$Be with correct asymptotic at large distance.
\end{abstract}

\maketitle

\section{Introduction}
In nuclear physics, the clustering phenomena have been intensively investigated by many experimental and theoretical studies~\cite{Descouvemont1989, Oertzen2006, Horiuchi1991}. One of the most important examples is the Hoyle state~\cite{Hoyle1954, Cook1957} which has clustered structure composed of three $\alpha$ particles, and plays an essential role for the synthesis of carbon in the universe~\cite{Fynbo2005}. For decades, the clustering in light stable nuclei has been studied by various experimental methods such as the resonant scattering, transfer reactions and the break-up reactions~\cite{Liu2020, Ma2021}. On the other hand, the clustering in neutron-rich nuclei is not known in as much detail as that of stable nuclei, due to the limited experimental techniques~\cite{Kanada1995, Kimura2016}.

Recently, the proton-induced $\alpha$ knockout reaction (p,p$\alpha$) has been established as a quantitative measure for the $\alpha$-cluster formation at nuclear surface~\cite{Yoshida2016, Wakasa2017, Yoshida2018, Yoshida2019, Yoshida2021}. The reaction has been used to measure the $\alpha$ clustering in Sn isotope chain~\cite{Tanaka2021} and revealed negative correlation between the $\alpha$-cluster formation and the neutron-skin thickness. This fascinating experiment has opened a gate to explore the $\alpha$ clustering in the light neutron-rich nuclei. In fact, the (p, p$\alpha$) experiment is under preparation to measure the $\alpha$ clustering in the Be and C isotopes. Combined with the data obtained from break-up reactions, it will provide comprehensive information on how the clustering in the ground and excited states evolves as a function of neutron number. Therefore, an accurate microscopic calculation is highly desired for the deeper understanding of the $\alpha$ clustering in the light neutron-rich nuclei.

In our previous work, we adopted the antisymmetrized molecular dynamics (AMD) framework to evaluate $\alpha$ formation probability in Be and C isotopes~\cite{Zhao2021, HeThesis}. The calculations showed the negative correlation between the $\alpha$ clustering and the neutron-skin thickness in the C isotopes, which is qualitatively consistent with what was observed in the Sn isotopes. However, from a quantitative point of view, those AMD results may underestimate the $\alpha$ formation probability due to the following reasons. Firstly, the Gogny D1S density functional in our previous study cannot describe light $s$- and $p$-shell nuclei accurately, because its parameters are optimized for medium- and heavy-mess nuclei~\cite{Gogny}. In particular, it does not reproduce the $\alpha$ and $^6$He decay threshold energies and radii of He isotopes, both of which strongly affect the cluster formation probability. Secondly, in our AMD+GCM calculation, we have used the quadrupole deformation parameter $\beta$ as the generator coordinate. However, this choice does not span the model space large enough to describe various cluster motions in nuclei, in particular, that at a large inter-cluster distance. In fact, it has been pointed out that an AMD+GCM calculation undershoots the $\alpha$ cluster formation probability of $^{48}$Ti in order of magnitude~\cite{Taniguchi2021}.

To overcome those defects, we perform alternative theoretical calculations with the following improvements. Firstly, we adopt the Volkov No.2 interaction~\cite{Volkov1965} for the nucleon-nucleon interaction which reproduces the radius of the $\alpha$ particle and the phase shift of the $\alpha$-$\alpha$ scattering. As for the spin-orbit interaction, we have employed the G3RS interaction. We have tuned the parameter set of these interactions to reproduce the threshold energies. We also note that the combination of those interactions has been adopted by many microscopic calculations to discuss the energy spectrum of light nuclei~\cite{Kanada2012, Itagaki2003}. Secondly, we employ the real-time evolution method (REM) to generate basis wave functions to describe various cluster motions. This method has successfully described the $\alpha$ clustering of Be, and C isotopes~\cite{Imai2019, Zhou2020, Zhao2022}. With these improvements, we have obtained a more reliable cluster formation probability which has the correct asymptotic at a large distance. As we a-priori assume the $\alpha+\alpha+x$n cluster structure of $^{10}$Be and $^{12}$Be, this work will give the upper limit of $\alpha$ cluster formation probability, whereas our previous study by AMD may give the lower limit.

This paper is organized as follows. In the next section, the theoretical framework of the generator coordinates method (GCM) with the real-time evolution method (REM) and the method to evaluate the reduced width amplitude (RWA) are briefly explained. In section \ref{sec:results}, we present the numerical results and discuss the differences in the results between the AMD and the REM. The final section summarizes this work.

\section{Theoretical Framework}
\label{sec:framework}

\subsection{The Hamiltonian and the wave function}
The Hamiltonian adopted in this study is given as
\begin{equation}
\hat{H}=\sum_{i=1}^A \hat{t}_i - \hat{T}_{c.m.} + \sum_{i<j}^A \hat{v}_N(\bm{r}_{ij}) + \sum_{i<j}^A \hat{v}_{C}(\bm{r}_{ij}) + \sum_{i<j}^A \hat{v}_{LS}(\bm{r}_{ij})~,
\end{equation}
where $\hat{t}_i$ and $\hat{T}_{c.m.}$ denote the kinetic energy operators of each nucleon and the center of mass, respectively. $\hat{v}_N$, $\hat{v}_C$, $\hat{v}_{LS}$ denote the effective nucleon-nucleon interaction, the Coulomb interaction and the spin-orbit interaction, respectively.

In this work, we use the Volkov No.2 interaction for the central nucleon-nucleon interaction~\cite{Volkov1965}, which is expressed as
\begin{equation}
\begin{split}
\hat{v}_N(\bm{r}_{ij})=&(W - M\hat{P}^\sigma \hat{P}^\tau + B\hat{P}^\sigma - H\hat{P}^\tau)\\
&\times [V_1\text{exp}(-\bm{r}_{ij}^2/c_1^2)+V_2\text{exp}(-\bm{r}_{ij}^2/c_2^2)] ~,
\end{split}
\end{equation}
where $W$, $M$, $B$ and $H$ denote the Wigner, Majorana, Bartlett, and Heisenberg exchanges, whose strengths are explained in the next section. The other parameters are, $V_1 = -60.65$ MeV, $V_2 = 61.14$ MeV, $c_1 = 1.80$ fm and $c_2 = 1.01$ fm. As for the spin-orbit interaction, we use the G3RS potential~\cite{Tamagaki1968, Yamaguchi1979}, 
\begin{equation}
\hat{v}_{LS}(\bm{r}_{ij}) = V_{ls}(e^{-d_1 \bm{r}_{ij}^2} - e^{-d_2 \bm{r}_{ij}^2})\hat{P}_{31}\hat{L}\cdot \hat{S}~.
\end{equation}
Here $\hat{P}_{31}$ projects the two-body system into triplet odd state, which can be expressed as $\hat{P}_{31}=\frac{1+\hat{P}^\sigma}{2}\cdot\frac{1+\hat{P}^\tau}{2}$. The Gaussian range parameters $d_1$ and $d_2$ are set to be $5.0$ fm$^{-2}$ and $2.778$ fm$^{-2}$, respectively. 

We approximate the He and Be isotopes as being composed of $\alpha$-clusters plus valence nucleons. Thus, the wave functions of these isotopes are written as 
\begin{equation}
\Phi(\bm{z}_{\alpha_1}...Z_1,Z_2...)=\mathcal{A}\{\Phi_\alpha(\bm{z}_{\alpha_1})...\phi(Z_1)\phi(Z_2)...\}~, 
\end{equation}
where $\Phi_\alpha(\bm{z}_{\alpha})$ is the antisymmetrized wave function of the $\alpha$-cluster with $(0s)^4$ configuration oriented at $\bm{z}_{\alpha}$. $\phi(Z)$ are the single-neutron wave functions. The single-particle wave function $\phi(\bm{r},Z)$ are expressed in a Gaussian form multiplied by the spin-isospin part $\chi\tau$ as 
\begin{equation}
\begin{split}
\phi(\bm{r},Z) =
 (\frac{2\nu}{\pi})^{3/4}&\text{exp}[-\nu(\bm{r}-\frac{\bm{z}}{\sqrt{\nu}})^2+\frac{1}{2}\bm{z}^2]\chi\tau~,\\
 &Z\equiv(\bm{z}, a, b)~.
\end{split} 
\end{equation}
Here $Z$ represents the time-dependent parameters of the wave function, which includes the three-dimensional coordinate $\bm{z}$ for the spatial part of the wave function as well as the spinor $a$ and $b$ for the spin part $\chi = a\ket{\uparrow}+b\ket{\downarrow}$. The isospin part is $\tau=\{\text{proton or neutron}\}$. The harmonic oscillator parameter is set to $b=\sqrt{1/(2\nu)}=1.46$ fm for both of $\alpha$ cluster and neutron wave functions, which is same with that used in Refs.~\cite{Itagaki2003,Furumoto2018}. 

\subsection{Real-time evolution method}

We use the real-time evolution method (REM)~\cite{Zhou2020, Zhao2022} to generate the basis wave functions which have various cluster configurations in the phase space. From the time-dependent variational principle:
\begin{equation}
\delta\int dt\frac{\bra{\Phi(\bm{z}_{\alpha_1}...Z_1,Z_2...)}i\hbar d/dt - \hat{H}\ket{\Phi(\bm{z}_{\alpha_1}...Z_1,Z_2...)}}{\braket{\Phi(\bm{z}_{\alpha_1}...Z_1,Z_2...)|\Phi(\bm{z}_{\alpha_1}...Z_1,Z_2...)}}~.
\end{equation}
We obtain the equation of the motion (EOM) for all the time-dependent parameters $Z(t)$ as\begin{equation}
\label{eq:EOM1}
i\hbar\sum_{j=\alpha_1,1,2}\sum_{\sigma=x,y,z,a} C_{i\rho j\sigma}\frac{dZ_{j\sigma}}{dt}=\frac{\partial \mathcal{H}_\text{int}}{\partial Z^*_{i\rho}}~,
\end{equation}

\begin{equation}
\label{eq:EOM2}
\mathcal{H}_\text{int}\equiv\frac{\bra{\Phi(\bm{z}_{\alpha_1}...Z_1,Z_2...)}\hat{H}\ket{\Phi(\bm{z}_{\alpha_1}...Z_1,Z_2...)}}{\langle\Phi(\bm{z}_{\alpha_1}...Z_1,Z_2...)|\Phi(\bm{z}_{\alpha_1}...Z_1,Z_2...)\rangle}~,
\end{equation}

\begin{equation}
\label{eq:EOM3}
C_{i\rho j\sigma}\equiv\frac{\partial^2 \text{ln}\langle\Phi(\bm{z}_{\alpha_1}...Z_1,Z_2...)|\Phi(\bm{z}_{\alpha_1}...Z_1,Z_2...)\rangle}{\partial Z^*_{i\rho}\partial Z_{j\sigma}}~.
\end{equation}
By solving this EOM starting from a certain wave function, a series of basis wave functions for generator coordinate method (GCM) is obtained.

Under the framework of GCM, the basis wave functions with different parameters $Z$ are superposed to describe the total wave function
\begin{equation}
\Psi^{J^\pi}_M = \int_0^{T_{max}}dt\sum_{K=-J}^J \hat{P}^{J^\pi}_{MK}f_K(t)\Phi(\bm{z}_{\alpha_1}(t)...,Z_{1}(t),Z_{2}(t)...)~,
\end{equation}
where $\hat{P}^{J^\pi}_{MK}$ is the parity and the angular momentum projector. This integral can be approximately discretized as
\begin{equation}
\Psi^{J^\pi}_M = \sum_{i,K} \hat{P}^{J^\pi}_{MK}f_{i,K}(t)\Phi_i~.
\end{equation}
The corresponding coefficients $f_{i,K}$ and the eigen-energy $E$ are obtained by solving the Hill-Wheeler equation.

\subsection{$\alpha$ reduced width amplitude}

To evaluate the degree of $\alpha$ clustering, we calculate the $\alpha$ reduced width amplitude (RWA) from the obtained wave functions, which is defined as the overlap amplitude between the A-body wave function of the mother nucleus $\Psi$ and the reference state composed of the clusters with mass numbers $C_1$ and $C_2$,
\begin{equation}
ay_l(a) = a\sqrt{\frac{A!}{(1+\delta_{C_1C_2})C_1!C_2!}}\langle \frac{\delta(r-a)}{r^2}\Psi_{C_1}\Psi_{C_2}Y_l(\hat{r})|\Psi\rangle~,
\end{equation}
where $\Psi_{C_1}$ and $\Psi_{C_2}$ are the ground state wave functions of the two clusters. This equation is calculated by using the Laplace expansion method~\cite{Chiba2017}. The $\alpha$ clustering may be evaluated by the $\alpha$ spectroscopic factor, which is defined as the squared integral of the RWA,
\begin{equation}
S_\alpha=\int_0^\infty r^2 dr~y_l^2(r)~.
\end{equation}
It is noted that $S_\alpha$ is not normalized unity because of the antisymmetrized effects between the clusters. In addition, we also introduce the root-mean-square radius of RWA,
\begin{equation}
R_{rwa}=\sqrt{\frac{\int_0^\infty r^4 dr~y_l^2(r)}{\int_0^\infty r^2 dr~y_l^2(r)}}~,
\end{equation}
which is a measure for the average distance between clusters.

\section{Results}
\label{sec:results}
\subsection{Structure properties of $^{10, 12}$Be and applied interaction parameters}

For the calculation of $^{10}$Be, we set the strengths of the central and spin-orbit interactions as $W=0.4$, $M=0.6$, $B=H=0.125$, and $V_{ls}=2000$ MeV. This parameter set has already been adopted in many calculations for this nucleus~\cite{Kanada2012, Itagaki2003}. The obtained numerical results are denoted as "REM (set1)" in Table~\ref{table:10Be}, where the numerical results are compared with those obtained by the AMD calculation with the Gogny D1S interaction.
\begin{table*}[htbp]
  \begin{center}
    \caption{The numerical results of the $0^+$ ground states of $^{10}$Be, $^6$He, and $^4$He. ``Exp.", ``AMD" and ``REM (set1)" denote the results from experimental data, AMD calculations, and REM calculations, respectively. Rc denotes the charge radius which is calculated from the point proton radius. $\Delta$E($\alpha$) represents the threshold energy to $^{6}$He+$^{4}$He.} \label{table:10Be}
    \vspace{2mm}
 \begin{tabular*}{14cm}{ @{\extracolsep{\fill}} l c c c}
    \hline
$^{10}$Be  &Exp. &AMD &REM\\
    \hline
$E$($^6$He)      &$-29.27$ MeV &$-33.04$ MeV &$-27.68$ MeV\\
$E$($^4$He)      &$-28.30$ MeV &$-29.68$ MeV &$-27.57$ MeV\\
$E$($^{10}$Be)  &$-64.98$ MeV &$-66.46$ MeV &$-62.46$ MeV\\
$R_c$($^{10}$Be)&$2.34$ fm       &$2.62$ fm        &$2.55$ fm\\
$\Delta E$($\alpha$) &$-7.41$ MeV   &$-3.75$ MeV   &$-7.21$ MeV\\
    \hline
  \end{tabular*}
  \end{center}
\end{table*} 
The $\alpha$-decay threshold energy is an important quantity for the discussion of the $\alpha$-cluster formation. We see that the AMD fails to reproduce it, while the REM yields a consistent result with the experimental data. The charge radius is another important quantity to discuss the $\alpha$ formation probability. The AMD results overestimate the charge radii of light nuclei. The REM also overestimates them but is smaller than the AMD result. In short, the REM with the Volkov No.2 interaction provides a better description of $\alpha$ threshold energy and the charge radius. We will discuss how these differences affect the RWA in the next subsection.

It has been known that the $^8$He+$^4$He and $^6$He+$^6$He configurations and even the $^7$He+$^5$He configuration play important role for the low-lying states of $^{12}$Be~\cite{Ito2012, Ito2004, Kimura2016, Oertzen2006}. The ground state of $^{12}$Be is dominated by the configuration with the breaking of $N=8$ magic number~\cite{Ito2012, Kanada2012, Shimoura2007}, and the $0_2^+$ state has $N=8$ closed shell structure. The numerical results are shown as "REM (set1)" in Table~\ref{table:12Be}. We found that REM does not reproduce many properties of $^{12}$Be if we apply the same interaction parameters as $^{10}$Be. 
\begin{table*}[htbp]
  \begin{center}
    \caption{The numerical results of the $0^+$ ground states of $^{12}$Be, $^8$He, and $^4$He. The meanings of the symbols are the same as in the previous table. Ex($0^+$) denotes the excitation energy of the first excited $0^+$ state. $\Delta$E($\alpha$) and $\Delta$E($^6$He) represent the threshold energies to $^8$He+$^4$He and $^6$He+$^6$He, respectively.} \label{table:12Be}
    \vspace{2mm}
 \begin{tabular*}{14cm}{ @{\extracolsep{\fill}} l c c c c}
    \hline
$^{12}$Be &Exp. &AMD &REM (set1) &REM (set2)\\
    \hline
$E$($^{12}$Be)  &$-68.65$ MeV &$-68.70$ MeV &$-60.89$ MeV &$-69.80$ MeV\\
$E_x$($0^+$)       &$2.25$ MeV    &$1.96$ MeV    &$5.54$ MeV    &$2.76$ MeV\\
$R_c$($^{12}$Be)&$2.50$ fm       &$2.82$ fm        &$2.47$ fm        &$2.45$ fm\\
$E$($^8$He)      &$-31.40$ MeV &$-33.65$ MeV &$-26.18$ MeV &$-35.64$ MeV\\
$E$($^6$He)      &$-29.27$ MeV &$-33.04$ MeV &$-27.68$ MeV &$-29.85$ MeV\\
$E$($^4$He)      &$-28.30$ MeV &$-29.68$ MeV &$-27.57$ MeV &$-27.57$ MeV\\
$\Delta E$($\alpha$)  &$-8.96$ MeV   &$-5.37$ MeV   &$-7.13$ MeV   &$-6.59$ MeV\\
$\Delta E$($^6$He)  &$-10.11$ MeV &$-2.63$ MeV   &$-5.53$ MeV   &$-10.11$ MeV\\
    \hline
  \end{tabular*}
  \end{center}
\end{table*} 
Furthermore, the order of $\alpha$+$^8$He and $^6$He+$^6$He channels, denoted by "$\Delta E$($\alpha$)" and "$\Delta E$($^6$He)", is inverted from the experimental data. We also find that the breaking of the magic number in the ground state can not be reproduced from the RWA results. Therefore, we slightly changed the Majorana parameters and the strength of the spin-orbit interaction to be $M=0.58$ ($W=1-M$) and $V_{ls}=2800$ to reproduce the following properties; the breaking of the magic number in the ground state, the $\alpha$+$^8$He and $^6$He+$^6$He threshold energies, and the excitation energy of the $0^+_2$ state. The results obtained by the fitted interaction are denoted by "REM (set2)" in Table~\ref{table:12Be}. This guarantees consistency with the observed data.

\subsection{RWA and Whittaker function}
\begin{figure*}[htbp]
  \begin{center}
   \includegraphics[width=0.7\hsize]{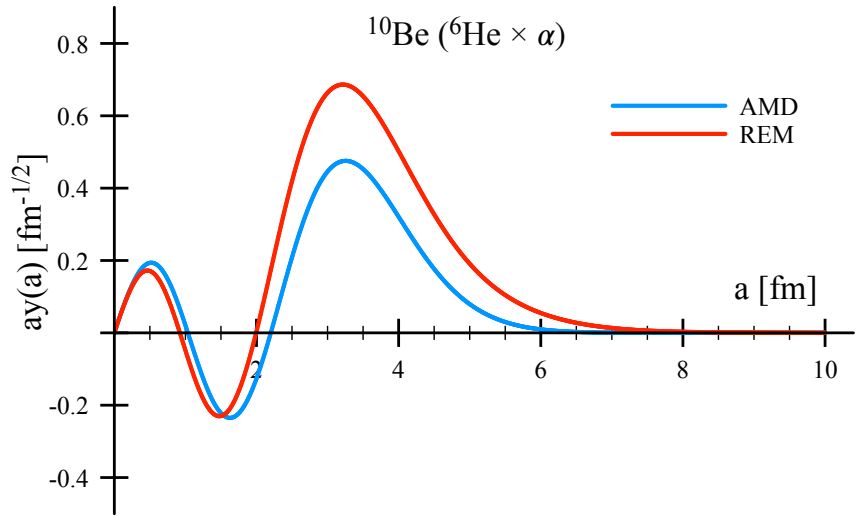}
  \caption{\label{fig:10berwa}The calculated RWA of the ground state of $^{10}$Be in the $^6$He+$^4$He channel. All the nuclei are in the ground state.} 
  \end{center}
\end{figure*}

Before we discuss the RWA, let us remind the structure of $^{10}$Be and $^{12}$Be. Two valence neutrons of $^{10}$Be occupy the $p$-shell, whereas two of four valence neutrons in $^{12}$Be are promoted to $sd$-shell across the $N=8$ shell gap. We then compare the RWA calculation results to see the similarities and differences in the $\alpha$ cluster formation probability obtained by AMD and REM frameworks. Fig.~\ref{fig:10berwa} shows the RWA of the ground state of $^{10}$Be for the $^6$He+$^4$He channel. 
It shows that the REM provides a much larger amplitude than the AMD calculation at the peak region, which indicates the larger possibility of the $\alpha$ formation in $^{10}$Be. This is due to the $\alpha$ cluster assumption in the REM wave function, while the AMD does not assume it. Therefore, the current REM result provides the upper limit of clustering. In addition to the difference in the amplitude, it can be found that the positions of the nodes of RWA from REM are shifted slightly to the inner part compared with the AMD calculations, which indicates the shorter spatial distribution of the $\alpha$ cluster. This is because that REM gives a smaller radius of $^{10}$Be ground state. Thus, REM yields a more enhanced but narrower distribution of $\alpha$ cluster.
\begin{table*}[htbp]
  \begin{center}
    \caption{The calculated r.m.s. radii of point neutron and proton distributions. All the units are in fm.} \label{table:radius}
    \vspace{2mm}
 \begin{tabular*}{14cm}{ @{\extracolsep{\fill}} l c c c c c c}
    \hline
         & \multicolumn{2}{c}{$^{10}$Be(g.s.)} & \multicolumn{2}{c}{$^{12}$Be(g.s.)} & \multicolumn{2}{c}{$^{12}$Be($0^+_2$)}\\
    \hline
          &AMD    &REM    &AMD     &REM (set2)     &AMD    &REM (set2)\\
    \hline
$r_n$ &$2.50$ &$2.53$    &$2.91$  &$2.50$    &$2.67$ &$2.52$\\
$r_p$ &$2.43$ &$2.35$    &$2.63$  &$2.23$    &$2.52$ &$2.21$\\
    \hline
  \end{tabular*}
  \end{center}
\end{table*} 

\begin{figure*}[htbp]
  \begin{center}
   \includegraphics[width=0.7\hsize]{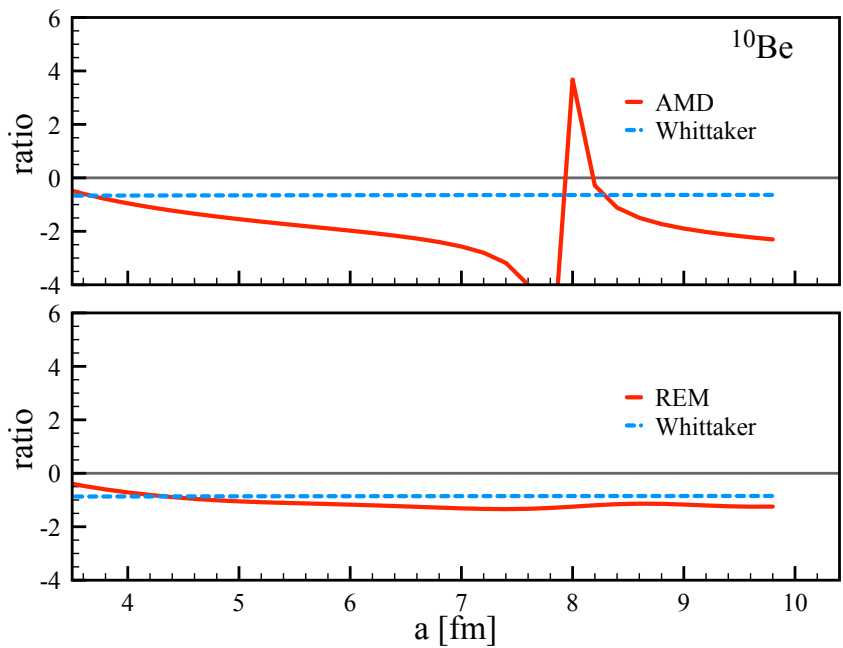}
  \caption{\label{fig:10bew}The comparison results between the calculated RWA and the Whittaker function for $^{10}$Be beyond $3.5$ fm. The vertical axis shows the ratio between the derivation of the functions as Eq.~\ref{eq:ratiow}.} 
  \end{center}
\end{figure*}

An advantage of REM over AMD is that it can describe the correct asymptotics of RWA. As explained in the appendix, the RWA should be identical to the Whittaker function at a large distance. We compare the logarithmic derivatives of the RWA and the Whittaker function in Fig.~\ref{fig:10bew}.
We can see that the RWA calculated from the AMD is inconsistent with the Whittaker function, whereas REM yields the correct asymptotics. From this result, we can easily obtain the asymptotic normalization constant (ANC), which is calculated as $5.3$.

\begin{figure*}[htbp]
  \begin{center}
   \includegraphics[width=0.7\hsize]{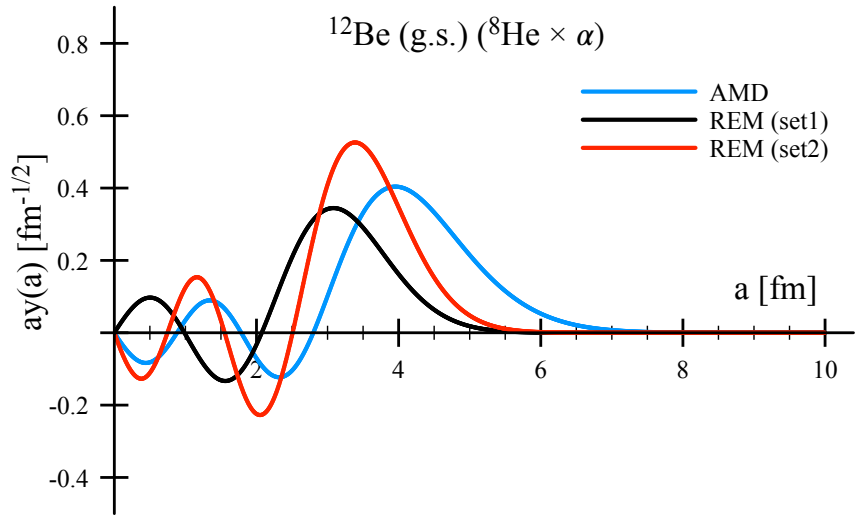}
  \caption{\label{fig:12berwa}The calculated RWA of $^{12}$Be in the $^8$He+$^4$He channel. All the nuclei are in the ground state.} 
  \end{center}
\end{figure*}

Unlike $^{10}$Be, the discussion on $^{12}$Be is complicated because of its more exotic structure. In Fig.~\ref{fig:12berwa}, we show the RWA of the ground state of $^{12}$Be calculated by using the original and modified interaction parameters.
Firstly, we note that the number of the nodes of RWA is different depending on the choice of the interaction parameter sets. This is due to the difference in the structure of the ground state. In the case of the original parameter set, all valence neutrons occupy the $p$-shell in contradiction to the experimental fact. On the other hand, in the case of the modified parameter set, the ground state is dominated by the $2\hbar\omega$ configuration, in which two valence neutrons are excited into the $sd$-shell across the $N=8$ shell gap. As a result, the RWA obtained by using the modified parameter set has an additional node. Secondly, we again see a larger amplitude at the peak value in the modified parameter set calculation than the AMD result, which can also be attributed to the cluster assumption node in the REM calculation. We also see that the RWA from the AMD calculation is much more spreading than the REM (set2) calculation. This is because of the much-overestimated radii by the AMD results.

\begin{figure*}[htbp]
  \begin{center}
   \includegraphics[width=0.7\hsize]{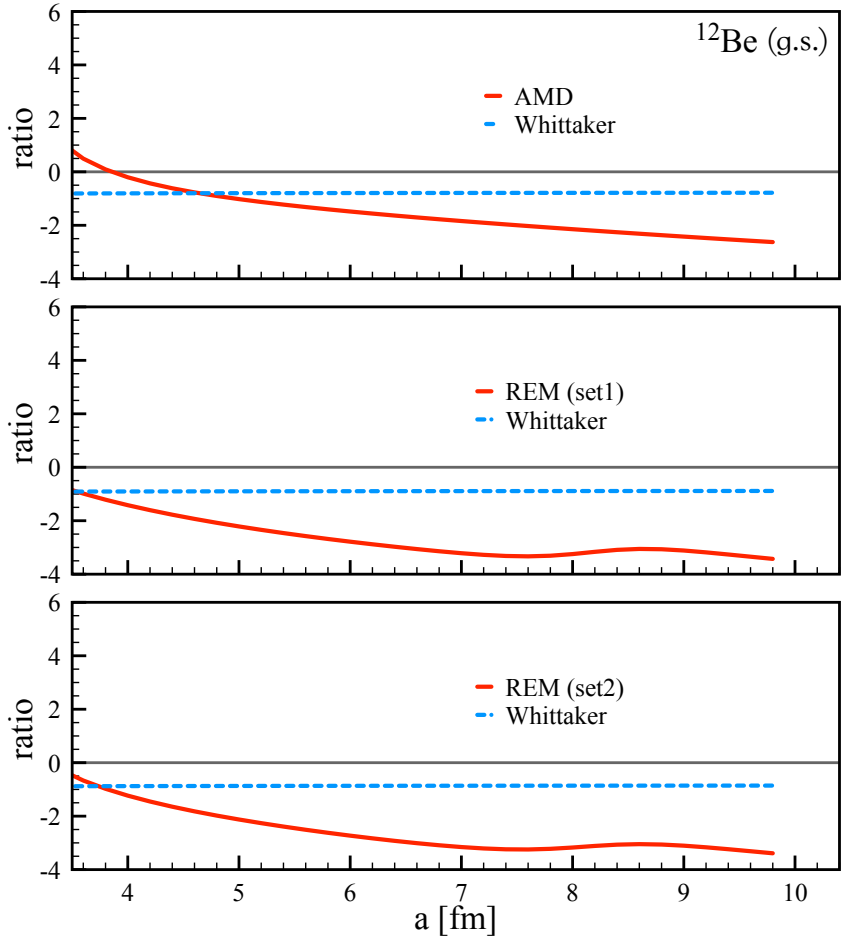}
  \caption{\label{fig:12bew}The comparison results between the calculated RWA and the Whittaker function for $^{12}$Be. The vertical axis shows the ratio between the derivation of the function as Eq.~\ref{eq:ratiow}.} 
  \end{center}
\end{figure*}

\begin{figure*}[htbp]
  \begin{center}
   \includegraphics[width=0.7\hsize]{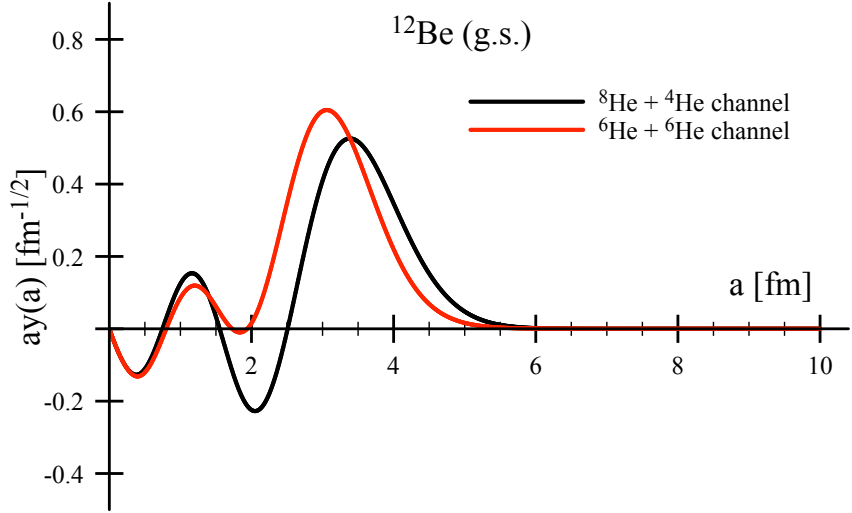}
  \caption{\label{fig:6he6he}The comparison of calculated RWA for $^{12}$Be between the $^8$He+$^4$He channel and the $^6$He+$^6$He channel. All the nuclei are in the ground state.} 
  \end{center}
\end{figure*}

The comparison between the RWA and the Whittaker function is shown in Fig.~\ref{fig:12bew}. The deviation from the Whittaker function are found for both of the AMD and REM (set2) calculations.
This result may be explained as follows. The Whittaker function is under the assumption of a two-body system. However, as already been discussed in many reference papers~\cite{Ito2012, Kanada2002}, the ground state of $^{12}$Be is an admixture of the $^{8}$He+$^4$He, $^{6}$He+$^6$He and $^5$He+$^7$He channels. This fact can also be found in Fig.~\ref{fig:6he6he}, where the compatible amplitude of RWAs for $^{12}$Be to $^{6}$He+$^{6}$He and $^{8}$He+$^{4}$He channels are shown.
Therefore, the ground state of $^{12}$Be can not be simply treated as a two-body system, which causes a departure from the Whittaker function.

\begin{figure*}[htbp]
  \begin{center}
   \includegraphics[width=0.7\hsize]{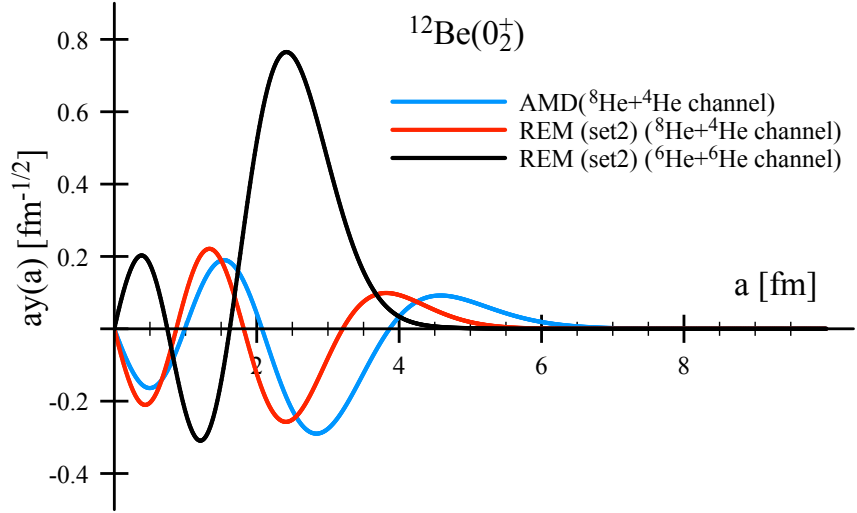}
  \caption{\label{fig:12be2rwa}The calculated RWA of the second $0^+$ state of $^{12}$Be in the $^8$He+$^4$He and $^6$He+$^6$He channels. All the nuclei are in the ground state.} 
  \end{center}
\end{figure*}

\begin{figure*}[htbp]
  \begin{center}
   \includegraphics[width=0.7\hsize]{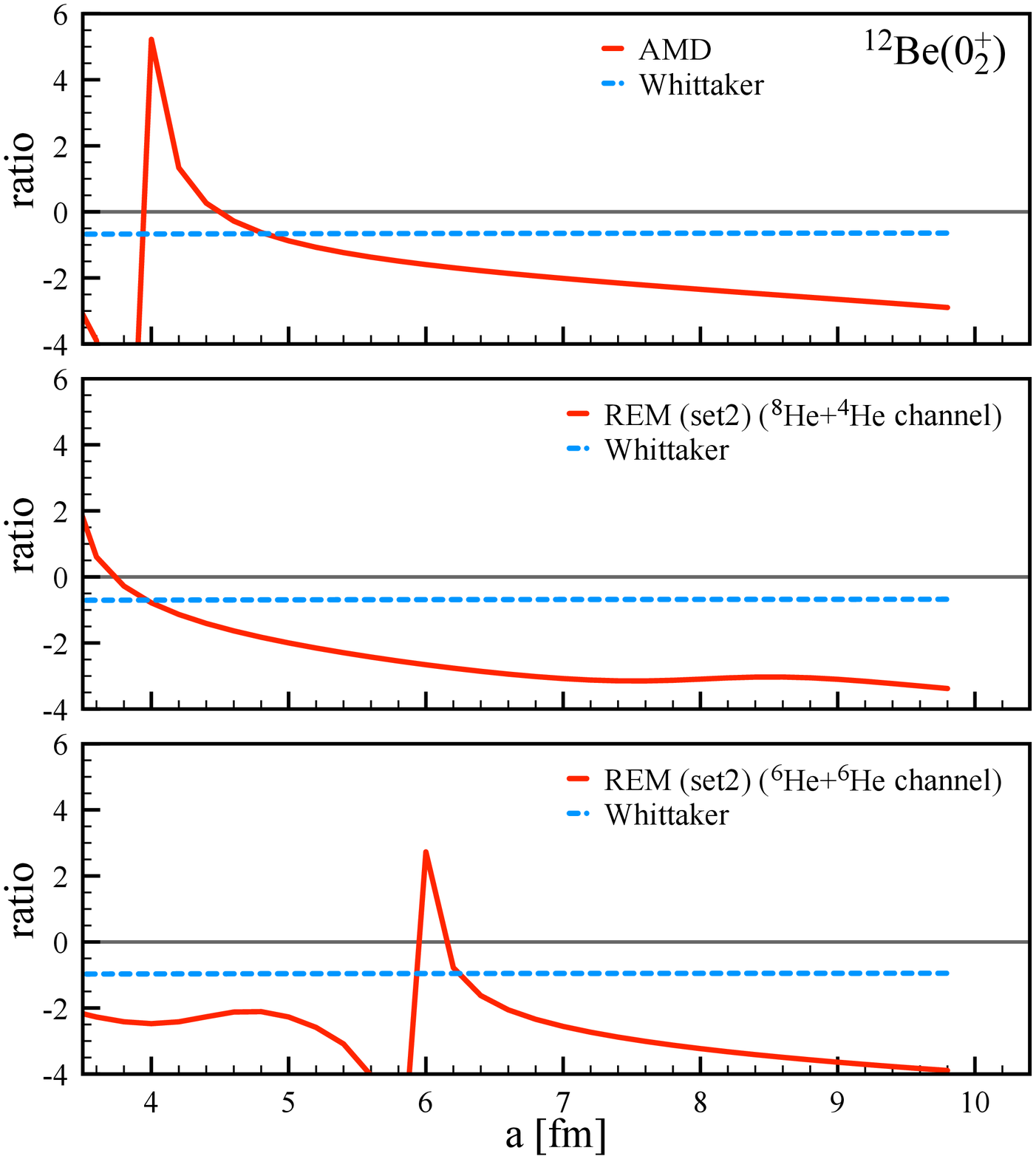}
  \caption{\label{fig:12be2w}The comparison results between the calculated RWA and the Whittaker function for $^{12}$Be. The vertical axis shows the ratio between the derivation of the function as Eq.~\ref{eq:ratiow}.} 
  \end{center}
\end{figure*}

In addition to the limitation of the two-body system assumption, the requirement that the cluster is only affected by the Coulomb interaction may also be violated by the RWA to cause the departure from the Whittaker function. As already been shown in Table~\ref{table:radius}, the r.m.s. radii of point neutron are much larger than the point proton for $^{12}$Be especially in the REM calculations. It suggests that the formed cluster in $^{12}$Be may be surrounded by the valence neutrons even in the outside region of the nucleus so that the cluster still feels the residual nuclear force. This residual nuclear force from the neutrons also leads to the departure from the Whittaker function. Another cause can be found from the RWA results of the $0_2^+$ state for $^{12}$Be shown in Fig.~\ref{fig:12be2rwa}.
It shows that the middle region of $^{12}$Be is dominated by the $^6$He+$^6$He channel while the outside region is dominated by the $^8$He+$^4$He channel, which should satisfy the assumption of two-body system. However, the amplitudes of RWA are only visible within about $5$ fm, where the nuclear force can not be neglected. This narrow distribution of RWA also cause the departure from the Whittaker function as shown in Fig.~\ref{fig:12be2w}.

Finally, we show the $\alpha$ spectroscopic factors and the r.m.s. radii of RWA in Table~\ref{table:sfactor}.
\begin{table*}[htbp]
  \begin{center}
    \caption{The calculated S-factor and r.m.s. radius of RWA.} \label{table:sfactor}
    \vspace{2mm}
 \begin{tabular*}{14cm}{ @{\extracolsep{\fill}} l c c c c}
    \hline
              & \multicolumn{2}{c}{$^{10}$Be} & \multicolumn{2}{c}{$^{12}$Be} \\
    \hline
              &AMD              &REM         &AMD            &REM (set2)\\
    \hline
$S_\alpha$ &$0.35$      &$0.77$     &$0.24$          &$0.33$\\
$R_{rwa}$ &$3.23$ fm  &$3.42$ fm &$4.07$ fm    &$3.36$ fm\\
    \hline
  \end{tabular*}
  \end{center}
\end{table*} 
It explicitly shows that the AMD result may provide a lower limit of the $\alpha$ clustering in the Be isotopes, while the REM procedure with the Volkov No.2 interaction may provide the upper limit. Both calculations show the suppression of the clustering with increasing neutron number, especially in REM calculation, which satisfies the discussion on the effect of the neutron-skin thickness in Ref.~\cite{Tanaka2021} and our previous work~\cite{Zhao2021}. In addition, AMD suggests a counter-intuitive enhanced r.m.s. radii of RWA, which describes a broader average region of $\alpha$ clustering even with the suppression by the neutron skin thickness. However, the REM calculations on the contrary can show a slightly suppressing in radii. These results again suggest that the REM procedure with Volkov No.2 interaction under the cluster model could be more proper for the description of the $\alpha$ clustering in the light nuclei.

\section{Summary}
\label{sec:summary}
The RWAs of $^{10}$Be and $^{12}$Be have been calculated with the REM procedure, where the cluster model and the Volkov No.2 interaction are adopted. Compared with the results from previous AMD calculations, the new results can reproduce the charge radii and the threshold energies. These quantities could be more important for the description of the $\alpha$ clustering. The RWA results from these frameworks show that the REM procedure may provide the upper limit of $\alpha$ clustering, while the AMD gives the lower limit. Besides, the suppression in r.m.s. radius of RWA by the neutron skin thickness can be correctly obtained by the REM calculations.

The Whittaker function is expected to be an important criterion for testing the calculation of cluster formation. The comparison with the Whittaker function shows that the current REM result of $^{10}$Be provides the correct asymptotic at large distance, which indicates more reliable description on the $\alpha$ cluster formation.

For the calculation of $^{12}$Be, the breaking of $N=8$ magic number is reproduced by slightly modifying the parameters of the interaction. The RWA results of $^{12}$Be suggest the mixing of $^{8}$He+$^4$He channel and $^{6}$He+$^6$He channel in the ground state. This complex structure causes the departure from the Whittaker function. In addition to this, the narrow distribution of the RWA and the influence of the valence neutrons may also affect the asymptotic of RWA. The behavior of RWA in the nuclei with exotic cluster structures or neutron skin should be further investigated in our future works.

% cSpell: disable 
\begin{acknowledgments}
The authors thank Dr. Z. H. Yang and Dr. K. Ogata for the fruitful discussions. This work was supported by National Natural Science Foundation of China [Grant Nos. 12147219, 11961141003], and JSPS KAKENHI [Grant Nos. 19K03859, 21H00113 and 22H01214]. Numerical calculations were performed in the server at the Theoretical Nuclear Physics Laboratory, Hokkaido University, and the Cluster-Computing Center of School of Science (C3S2) at Huzhou University.
\end{acknowledgments}

\begin{appendix}
\section{Whittaker function}

The RWA can be treated as the wave function of the $\alpha$ cluster in the mother nucleus. Therefore, it should be the solution of the time-independent Schr{\"o}dinger equation:
\begin{equation}
\label{eq:Sch}
-\frac{\hbar^2}{2m}\nabla^2\psi+V(r)\psi=E\psi~.
\end{equation}
In the case of two-body system, $r$ is the distance between two particles and $m$ is the reduced mass. $V(r)$ represents the potential between two particles and the eigenvalue $E$ will be the threshold energy between them.

Considering the situation when the formed cluster is too far away from the residue nucleus, only the Coulomb potential is present, then the solutions of Eq.~(\ref{eq:Sch}) are the Whittaker functions~\cite{Descouvemont2010, Brune2020}:
\begin{equation}
\begin{split}
&u_1=M_{-\eta,l+\frac{1}{2}}(2kr)\\
&u_2=W_{-\eta,l+\frac{1}{2}}(2kr)
\end{split}
\end{equation}
where $l$ is relative angular momentum between two particles and other parameters are summarised as
\begin{equation}
\begin{split}
{\rm reduced~mass:}~&m=m_1 m_2/(m_1+m_2)\\
{\rm wave~number:}~&k=\sqrt{-2mE/\hbar^2}\\
{\rm dimensionless~Sommerfeld~parameter:}~&\eta=Z_1Z_2e^2m/4\pi\epsilon_0\hbar^2k~.
\end{split}
\end{equation}
Among these two solutions, we only adopt the Whittaker W function for the physical meaning.

In this sense, if the RWA calculated from a framework is accurate enough, it should be identical to the Whittaker function starting from a large distance $a$ with an normalization factor $C$ as
\begin{equation}
\label{eq:whittaker}
ry_l(r)=CW(r)~\text{for}~r>a~.
\end{equation}
This factor C is called Asymptotic Normalization Constant (ANC), which is of importance for the study of the astrophysical reactions. In addition, if C is really a constant in the calculation, the derivation of Eq.~\ref{eq:whittaker} should still hold as
\begin{equation}
\label{eq:dwhittaker}
[ry_l(r)]'=CW'(r)~\text{for}~r>a~.
\end{equation}
Therefore, by comparing the ratio between Eq.~\ref{eq:whittaker} and Eq.~\ref{eq:dwhittaker} as following equations:
\begin{equation}
\label{eq:ratiow}
\frac{[ry_l(r)]'}{ry_l(r)}~,~~\frac{W'(r)}{W(r)}~.
\end{equation}
the existence of the constant C can be confirmed if these two ratios are identical. The comparison between RWA and Whittaker equation can be treated as an important criterion of the accuracy of the calculation. In this work, we will compare the RWA obtained from AMD and REM calculations with the Whittaker function to discuss the model dependence of the description of cluster formation.

\end{appendix}

\end{document}